\documentclass[12pt,english]{article}
\usepackage[T1]{fontenc}
\usepackage[latin9]{inputenc}
\usepackage{array}
\usepackage{multirow}
\usepackage{amsmath}
\usepackage{amssymb}
\usepackage{graphicx}

\makeatletter

\providecommand{\tabularnewline}{\\}

\@ifundefined{definecolor}
 {\usepackage{color}}{}

\makeatother

\usepackage{babel}
\begin{document}

\title{\textbf{Effect of Massive Neutrino on}\\
\textbf{ Large Scale Structures }}

\author{P. R. Dhungel, S. K. Sharma and U. Khanal \\
 Central Department of Physics,\\
 Tribhuvan University, Kirtipur, Nepal}

\date{\today}
\maketitle
\begin{abstract}
Jeans mass calculated with different combination of parameters involved
has shown interesting variation with remarkable shifting of position
of peak value from x = m/T = 0.5 to 5.5. The standard deviation is
2.217. In particular, using the harmonic mean square velocity, shifts
the peak Jeans mass to x = m/T $\sim$ 2, which is remarkably less
than previously reported value of 4.2. Different scales of neutrino
structures including virialized moments have also been compared.
\end{abstract}

\section{\protect\bigskip{}
 Introduction}

The enigmatic neutrino has generated great deal of interest in cosmology
for a long time. Regarding the number of neutrino species e.g., calculation
of primordial nuclear abundances in cosmology had arrived at the number
of neutrino species $N_{\nu}<4$ \cite{steigman} long before it was
experimentally established at CERN\cite{nuncern}; the current value
is $N^{\nu}=2.994\pm0.013$ \cite{yao}. Similar calculations set
the baryon to photon ratio at $\sim4.7\times10^{-10}$, which in turn
has determined the present baryon density in the Universe to be $\Omega_{B}h^{2}=0.0223$
, where $\Omega_{B}=\rho_{B}/\rho_{c}$ is the present baryon density
in units of the critical density, and h is the present value of the
Hubble parameter in units of $100\, km/s/Mpc$, that is expected to
be $h\sim0.72(3)$\cite{pdg}. Writing the present neutrino density
as $\Omega_{\nu}h^{2}=m_{eV}/31$, where $m_{eV}$ is the average
rest-mass of the three species neutrino in $eV$, it is easily seen
that the\ $31\, eV$ neutrino will close the Universe; this closure
mass is referred to as the Cowsik-McClelland bound\cite{Cowsik},
although in their paper they had used a smaller value of $\Omega$
and four component neutrino to arrive at $8\, eV$. Tremain and Gunn
\cite{TG} set the limit that a neutrino of $m_{\nu}<1MeV$ cannot
reside in the halo of galaxy to contribute significantly to the dark
matter. Bond et al \cite{Bond} calculated the maximum Jeans mass
for structures of neutrino to be $M_{\nu m}\approx1.2\times10^{17}m_{eV}^{-2}M_{\bigodot}$
, where $m_{eV}$ is the neutrino mass in eV.

As recent results show that the mass difference between neutrino species
is very slight, $\Delta m^{2}\sim10^{-5}\, eV^{2}$, in this paper,
we consider three neutrino species of similar mass. Comparing the
baryon and neutrino densities, we see that even a $0.7\, eV$ neutrino
will have dominated over baryon by now. Rather than whether the neutrino
is massive, the question at present is how massive it is; again cosmology
provides the most stringent limit $\sum m_{\nu}<0.6\, eV$ \cite{Spergel,7YrWMAP10}.
So the neutrino appears as a very important component of the Universe.
Consequently, it should have a strong bearing on structure formation.
It decouples at very early times, and then evolves as a totally independent
component that interacts only gravitationally. Indeed the filamentary
structures, sheets, walls and voids as exposed by various surveys
point towards dissipationless collapse of clouds of particles like
neutrinos into Zeldovich pancakes\cite{Zeld} at some stage of evolution.
Although the cosmological neutrino has been studied extensively, we
feel that our analysis gives some further insight on structure formation,
particularly at what temperature the structures are more likely to
form and spectrum of the size / mass of the structures at various
temperatures.

The momentum of a freely moving particle in the Friedmann-Robertson-Walker
spacetime is redshifted by the expansion, i.e., the comoving momentum
$pa=y$ remains constant, where $p$ is the momentum and $a$ the
scale factor. Using $v=p/E$ for the velocity and the Einstein energy-momentum
relation, $a^{2}E^{2}=p^{2}a^{2}+m^{2}a^{2}=y^{2}+x^{2}$, we see
that $Eva$ remains constant during expansion, and as the number density
scales as $a^{-3}$, $\rho va^{4}$ also remains constant where $\rho$
is the density. As the light neutrinos decouple at the very high temperature
of $T\approx1MeV$ while still extremely relativistic (ER), they are
essentially in free fall since then. So their number density is always
distributed as

\begin{equation}
dn=\frac{g}{2\pi^{2}}\left(\frac{m}{x}\right)^{3}\frac{y^{2}}{e^{y}+1}dy\label{1.1}
\end{equation}
where $g$ is the number of spin degeneracy (six, for the three $\nu-\overline{\nu}$
pairs), and we have used the fact that $T\sim1/a$ ; Planck units
in which $G=c=k_{B}=\hbar=1$ are used throughout this work. As $E\sim p>>m$
in the extreme relativistic (ER) regime, Eq.(\ref{1.1}) represents
the Fermi-Dirac distribution. In the non-relativistic (NR) case however,
$E\sim p^{2}/2m+m$, and Eq. (\ref{1.1}) \ is no more a Fermi-Dirac
distribution.

\begin{figure}[h]

\centering{}\scalebox{.9}{\includegraphics{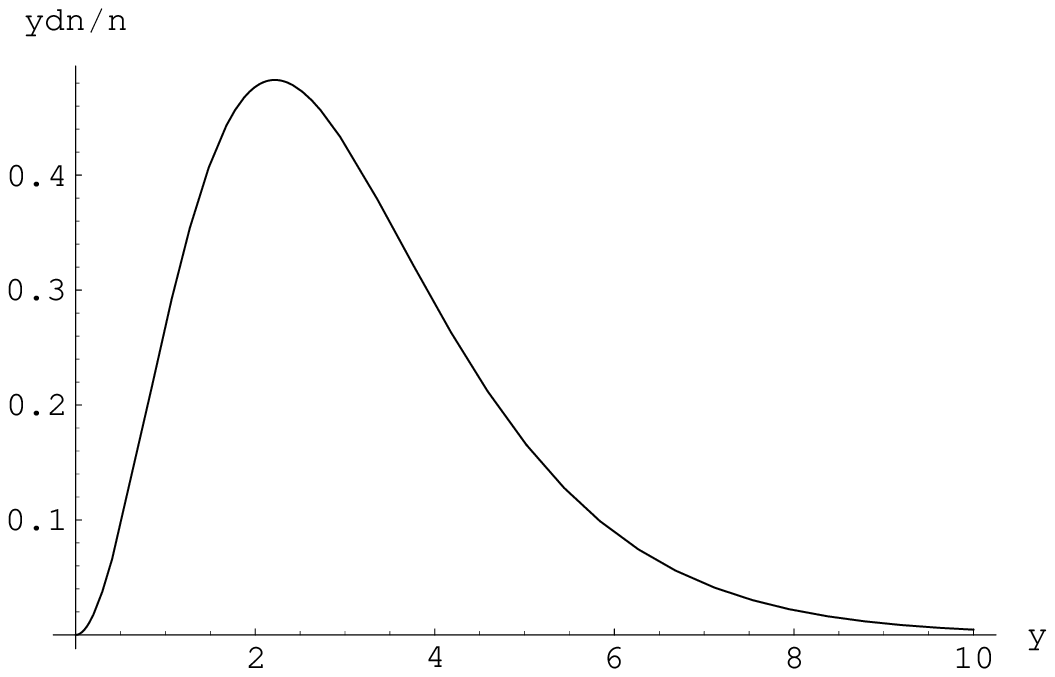}}
\caption{Momentum distribution of the free cosmic neutrino as a
function of $y=pa=p/T_{\nu}$. The distribution is maximum at
$y_{max}$ 3.131, and other characteristic values are
$y_{rms}=<y^{2}>^{1/2}=3.597$ and
$y_{rhms}=<1/y^{2}>^{-1/2}=1.613$. Such momentum distribution give
rise to a spectrum of neutrino structure
scales.\label{fig:momtmdist}}
\end{figure}

Integrating dn over $y$ from $0$ to $\infty$ gives the number density\\
 $n=\frac{g}{2\pi^{2}}\Gamma(3)\eta(3)\left(\frac{m}{x}\right)^{3}$
where $\Gamma$ and $\eta$ are the gamma and eta functions respectively\cite{Abram};
also, $\eta(n)=\left(1-2^{1-n}\right)\zeta(n)$. Thus we can write
down the expectation value of any regular function as $<f(y)>=\frac{1}{\Gamma(3)\eta(3)}\int_{0}^{\infty}dy\frac{y^{2}}{e^{y}+1}f(y)$.
\bigskip{}
 The width of the momentum distribution shown in Fig. ~\ref{fig:momtmdist}
is characterized by $y_{max}=3.131$ where the distribution is maximum.
Some other characteristic values that we will use are the mean $y_{mean}=<y>=\frac{\Gamma(4)\eta(4)}{\Gamma(3)\eta(3)}=\frac{7\zeta(4)}{2\zeta(3)}=3.151$,
the root-mean square $y_{rms}=<y^{2}>^{1/2}=\sqrt{\frac{15\zeta(5)}{\zeta(3)}}=3.597$,the
harmonic mean $y_{hm}=<y^{-1}>^{-1}=\frac{3\zeta(3)}{\zeta(2)}=2.192$,
and the root harmonic mean square $y_{rhms}=<y^{-2}>^{-1/2}=\sqrt{\frac{\Gamma(3)\eta(3)}{\Gamma(1)\eta(1)}}=\sqrt{\frac{3\zeta(3)}{2\ln(2)}}=1.613$
are also useful to describe the width of the distribution.

\begin{figure}[h]

\centering{}\scalebox{.9}{\includegraphics{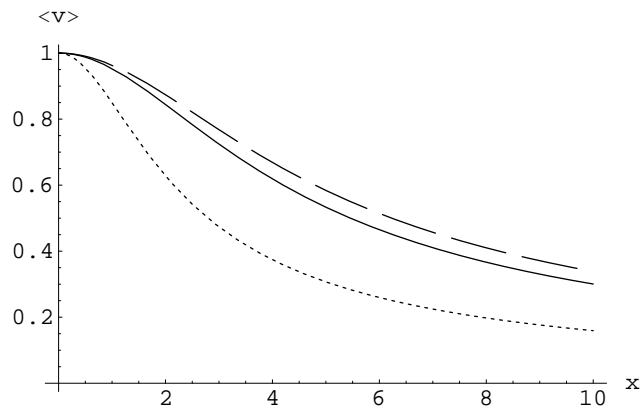}} \caption{The
average speed of the massive neutrino as a function of x. The
solid curve is $v_{mean}$, the dashed curve is $v_{rms}$ and the
dotted curve is $v_{rhms}$. In the extreme relativistic regime as
$x\rightarrow0$, it is seen that $v_{mean}\rightarrow v_{rhms}$;
towards the non-relativistic end it is closer to
$v_{rms}$.\label{fig:averagevel}}
\end{figure}

The particle speed is $v=p/E=y/\sqrt{y^{2}+x^{2}}$. In the ER case
as $y>>x$, $v(y)\rightarrow1-\frac{x^{2}}{2\, y^{2}}+.....$\ giving
$v_{mean}\rightarrow1-\frac{1}{2\,}\left(\frac{x}{y_{rhms}}\right)^{2}\rightarrow v_{rhms}$;
also, $v_{hm}=<1/v>^{-1}\rightarrow1-\frac{1}{2\,}\left(\frac{x}{y_{rhms}}\right)^{2}\rightarrow v_{rhms}$.
Three averages of v are shown in Fig. \ref{fig:averagevel}. Thus,
we see that $v_{rhms}$ is the more representative speed in the ER
regime. These ideas can be used for a bosonic system as well by replacing
$\eta$ with $\zeta$ . But $\zeta(1)=\infty$, so $v_{rhms}(boson)=0$,
indicative of the liability of bosons to condense into a zero momentum
state. Although a zero momentum state is strictly forbidden for a
fermionic system, as $\eta\left(1\right)=\ln\left(2\right)$, a low
momentum condensate characterized by the harmonic mean values is possible.
The rms energy is given by $a^{2}E_{rms}^{2}=y_{rms}^{2}+x^{2}$,
and another characteristic value of the energy is the velocity averaged
momentum $a^{2}E_{v}^{2}=<y^{2}><1/v^{2}>=y_{rms}^{2}(1+x^{2}/y_{rmhs}^{2})$
. Obviously, as $a\rightarrow0,\, E\rightarrow\infty$.

Any macroscopic quantity that depends on the momentum distribution
will be sensitive to the averaging process. This will be even more
so in the relativistic regime. In this paper, we look into such effects
and investigate the possible distribution of the sizes of neutrino
structures. In particular, we would like to determine whether smaller
neutrino structures could have formed in the very early Universe.
In the next Section, we apply these ideas to the gravitating neutrino
spheres, and compare the distributions of Keplerian, Virial and Jeans
and the free streaming scales.

\section{\protect\bigskip{}
 Neutrino Jeans Mass\label{sec:-Neutrino-Jeans-mass}}

The Jeans mass that is contained within the Jeans radius, $M_{J}=\frac{4\pi R_{J}^{3}\rho_{\nu}}{3}=\frac{1}{2}(\frac{R_{J}^{3}}{R_{\nu}^{2}})$,
has the momentum dependence \cite{Prem},

\begin{equation}
\frac{M_{J}}{C}=\frac{x^{2}y^{3}}{(y^{2}+x^{2})^{7/4}},\label{jeansmass}
\end{equation}
 where the constant $C=\frac{\pi^{7/2}m_{Pl}^{3}}{6\sqrt{g\eta(3)m^{2}}}=\frac{1}{m_{eV}^{2}\sqrt{g}}\times15\times10^{18}M_{\bigodot}$,
$m_{Pl}$ is the Planck mass, x = ma = m/T and y = pa = p/T . Any
gas cloud of radius greater than $R_{J}$ will contain a mass greater
than $M_{J}$, in which case gravitation can overcome the free streaming
motion of the neutrinos to produce a collapse. This analysis can be
used for any or all the components of the Universe, and here we will
apply it to the neutrino. These characteristic length and mass scales,
being dependent on the momentum distribution, are sensitive to the
averaging process.

\begin{figure}[h]

\centering{}\scalebox{.9}{\includegraphics{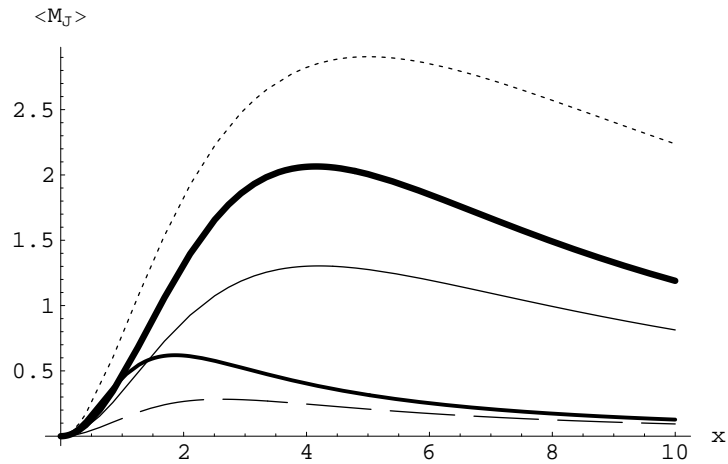}}
\caption{The distribution of Jeans Mass. The dashed curve
represents the value calculated by Bond, \textit{et al}, i.e.,
$<M_{J}>_{1}$ The dotted curve is $<M_{J}>_{2}$, the thick solid
curve is $<M_{J}>_{3}$, moderately thick curve is $<M_{J}>_{4}$
and the normal curve is $<M_{J}>_{5}$. Two distinct groups of
Jeans mass are seen, one that peaks at $x\sim4$ to a value of
$\sim4$, and another that peaks at $x\sim2$ to little less than 1.
The first group appears to be related to the rms speed and the
second to the rhms speed. As pointed out in the text, at the
relativistic regime it is the rhms speed that is more
representative. So the smaller second group of structures can
start to form with smaller mass at earlier times.\label{fig:jms}}
\end{figure}

J.R. Bond \textit{et al}\cite{Bond} calculated $M_{J}$ by comparing
the pressure and the density as $<M_{J}>_{1}\sim\langle\rho v^{2}\rangle^{3/2}/\langle\rho\rangle^{2}$,
which was found to peak at $x_{1}=4.2$. This expression may be appropriate
in the gravitational collapse of a mass of a gas against its internal
pressure, but the neutrinos are free-streaming since decoupling. Thinking
that $k_{J}$ is the fundamental quantity that determines the Jeans
scale, we calculated $<M_{J}>_{2}\sim\langle\rho v^{2}\rangle^{3/2}/\langle k_{J}\rangle^{3}$
and found the peak to be at $x_{1}=2.5$. We can also find the average
of the Jeans mass as $<M_{J}>_{3}=<M_{J}(y)>$ with peak at 5.0, and
represent some characteristic values analytically by $<M_{J}>_{4}=<M_{J}(y_{rms})>$
with peak at 4.2, and $<M_{J}>_{5}=<M_{J}(y_{rhms})>$ with peak at
1.9. Plot of these values are displayed in Fig. ~\ref{fig:jms}.
The values of x at which these $M_{J}$'s peak are given in the table
~\ref{tab:jms}. The results show quite a wide variations in the
Jeans mass ranging across a magnitude. Also the time when these structures
can form, determined by x, varies from 1.9 to 5. In essence two groups
of neutrino structures could appear: the smaller one at $x\sim2$
when the neutrino are still quite relativistic with $v_{rhms}\sim0.6$,
and the ten times larger one at $x\sim5$ when the neutrinos have
become non-relativistic with $v_{rhms}\sim0.3$.

\begin{table*}
\begin{centering}
\caption{\label{tab:jms} The value $x_{p}$ at which the different means of
neutrino Jeans mass peak and also the peak value. As the maximum of
the $x_{p}$'s is 5, the different $<M_{J}>$'s are caculated at this
value and the standard deviation is determined.}

\par\end{centering}

\centering{}%
\begin{tabular}{|c|c|c|c|c|}
\hline
Method  & $x_{p}$  & $M_{J}(x_{p})$  & $M_{J}(x=5)$  & Standard Deviation \tabularnewline
\hline
$<M_{J}>_{1}$  & 4.2  & 1.75  & 1.71  & \tabularnewline
\hline
$<M_{J}>_{2}$  & 2.1  & 0.73  & 0.36  & \tabularnewline
\hline
$<M_{J}>_{3}$  & 5.0  & 1.61  & 1.61  & 0.832 \tabularnewline
\hline
$<M_{J}>_{4}$  & 4.2  & 2.06  & 2.01  & \tabularnewline
\hline
$<M_{J}>_{5}$  & 1.9  & 0.41  & 0.21  & \tabularnewline
\hline
\end{tabular}
\end{table*}

\section{\protect\bigskip{}
 Neutrino structure scales}

To investigate the scales of neutrino structures, let us rewrite the
Friedmann equation in terms of the Hubble radius $R_{H}=1/H$ in the
following form,
\begin{equation}
\frac{R_{H_{0}}^{2}}{R_{H}^{2}}=\frac{H^{2}}{H_{0}^{2}}=\frac{R_{H_{0}}^{2}}{\xi^{2}}\left(\frac{d\xi}{dt}\right)=\frac{\rho}{\rho_{c0}}+\frac{1-\Omega}{\xi^{2}}=\frac{f(\xi)}{\xi^{4}}.\label{2.0}
\end{equation}
Here the subscript $0$ represents the respective values at some reference
time $t_{0}$ and $\rho_{c}=\frac{3H^{2}}{8\pi}$ is the critical
density; $\rho$ is the total energy density contributed by matter
and radiation, while $1-\Omega$ may be considered to be the energy
density due to curvature, and in particular, $f(\xi)=\Omega_{\Lambda0}\xi^{4}+(1-\Omega_{0})\xi^{2}+\Omega_{d0}\xi+\Omega_{\gamma0}+\Omega_{\nu rest0}\sqrt{(\xi^{2}+(y/x_{0}))}$
...., where $\xi=x/x_{0}=a/a_{0}=T_{0}/T,\Omega_{\Lambda,d,\gamma,\nu}$
are the respective contributions to the total density at $t_{0}$
(in units of critical density) by the cosmological constant, pressureless
dust, photon, neutrino, etc. In the way we have written the neutrino
density, we need not make the distinction between radiation and matter
domination as its density scales appropriately in the respective regime

$\rho_{\nu}/\rho_{c0}\rightarrow\begin{cases}
\Omega_{\nu0}/\xi^{4}, & \xi\rightarrow0,\,\,\,\,\,\,(ER)\\
\Omega_{\nu\, rest\,0}/\xi^{3}, & \xi\rightarrow0,\,\,\,\,\,\,(NR).
\end{cases}$

The ratio of neutrino and photon density is

\begin{equation}
\frac{\Omega_{\nu}}{\Omega_{\gamma}}=\frac{\rho_{\nu}}{\rho_{\gamma}}=\left(\frac{T_{\nu}}{T_{\gamma}}\right)^{4}\frac{3\zeta(3)}{4\zeta(4)}x\langle\sqrt{1+(y/x)^{2}}\rangle=\frac{\Omega_{\nu.rest}}{\Omega_{\gamma}}\langle\sqrt{1+(y/x)^{2}}\rangle.\label{eq:nuphotonratio}
\end{equation}

The temperature ratio $T_{\nu}/T_{\gamma}$ is unity at very high
temperature and is $(4/11)^{1/3}$ at temperatures below the neutrino
decoupling and $e^{+}-e^{-}$ annihilation, i.e., $T_{\nu}<1MeV$;
the density ratio rises linearly in the NR region. The temperature
of photon-neutrino equality can be determined by setting Eq.~(\eqref{eq:nuphotonratio})\,equal
to unity, whence one finds $x_{eq}=3.179$.

The most important length scale in cosmological context is the Hubble
radius; the ratio of the comoving Hubble radii at equality and at
arbitrary time is given by

\begin{equation}
\left(\frac{R_{H_{eq}}/a_{0}}{R_{H}/a}\right)^{2}=\left(\frac{\xi R_{H_{eq}}}{R_{H}}\right)^{2}=\frac{f_{\nu}(\xi)}{\xi^{2}}.\label{eq:rhr02rh}
\end{equation}

A length scale relevant to a gravitationally bound gaseous system
is the radius where the escape speed becomes equal to the average
random speed of the gas particles viz., $v^{2}=2M/R_{esc}=\frac{8\pi}{3}\rho R_{esc}^{2}$,
or $\frac{1}{R_{esc}^{2}}=\frac{8\pi\rho}{3v^{2}}$. As the smooth
background of photons will not contribute to the potential well, $\rho=\rho_{\nu}$.
We can use the expression for critical density to write the ratio
of the comoving Hubble and escape radii in a suggestive form that
looks similar to Eq. ~\eqref{eq:rhr02rh}:

\begin{equation}
\left(\frac{\xi R_{H_{0}}}{R_{esc}}\right)^{2}=\frac{f_{esc}(\xi)}{\xi^{2}},\label{eq:xirh02resc}
\end{equation}

where\\

\begin{equation}
f_{esc}(\xi)=\Omega_{\nu rest0}\langle\sqrt{\xi^{2}+(y/x_{0})^{2}}\rangle/v^{2}.\label{eq:fesc}
\end{equation}

We can also define the Keplerian radius $R_{K}$ by$\frac{1}{R_{K}^{2}}=\frac{4\pi\rho}{3v_{rot}^{2}}=\frac{1}{2R_{esc}^{2}}$
with the understanding that $v$ stands for the rotational speed;
obviously, any particle orbiting with $v_{rot}>v_{esc}$ will escape.
Similarly, the virial radius given by $2<v^{2}>=M<1/R_{V}>$ leads
to $\frac{1}{R_{V}^{2}}=\frac{4\pi\rho}{6v^{2}}=\frac{1}{4R_{esc}^{2}}$.
Another such scale, the Jeans radius is given by $\frac{1}{R_{J}^{2}}=\frac{4\rho}{\pi v^{2}}=\frac{3}{2\pi^{2}R_{esc}^{2}}$
. All these length scales are some numerical multiples of $R_{esc}$.

\begin{figure}[h]

\centering{}\scalebox{.9}{\includegraphics{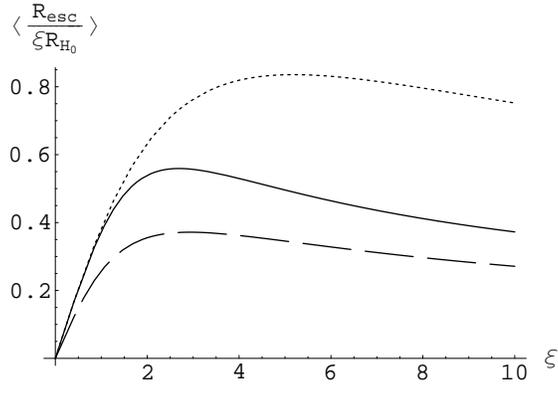}}
\caption{The ratio of $R_{esc}/\xi R_{H}$ for the three averages.
\label{fig:resc2xirh1}}
\end{figure}

In any spectral analysis, it is the wavenumber $k\sim1/R$ that is
fundamental, and similarly, in the Jeans analysis, the Jeans wavenumber
$k_{J}$ should be averaged over all momenta. We have seen that $R_{esc}$
is the underlying quantity, so let us look into the behaviour of $1/R_{esc}^{2}$.
There are a number of ways of averaging Eq. ~(\ref{eq:xirh02resc}),
as discussed in the previous section:

\begin{subequations}
\begin{align}
\left(\frac{\xi R_{H_{0}}}{R_{esc}}\right)_{1}^{2} & =\frac{\Omega_{\nu\, rest\,0}}{\xi^{2}}\left\langle \frac{(\xi^{2}+(y/x_{0})^{2})^{3/2}}{(y/x_{0})^{2}}\right\rangle \,\,\,\,\,\,\,\mathtt{from}\,\,\left\langle \frac{\rho}{v^{2}}\right\rangle \\
\left(\frac{\xi R_{H_{0}}}{R_{esc}}\right)_{2}^{2} & =\frac{\Omega_{\nu\, rest\,0}}{\xi^{2}}\left\langle \sqrt{\xi^{2}+(y/x_{0})^{2}}\right\rangle \left\langle \frac{\xi^{2}+(y/x_{0})^{2}}{(y/x_{0})^{2}}\right\rangle \,\mathtt{from}\langle\rho\rangle\left\langle \frac{1}{v^{2}}\right\rangle \\
\left(\frac{\xi R_{H_{0}}}{R_{esc}}\right)_{3}^{2} & =\frac{\Omega_{\nu\, rest\,0}}{\xi^{2}}\frac{\langle\sqrt{\xi^{2}+(y/x_{0})^{2}}\rangle^{2}}{\langle(y/x_{0})^{2}/\sqrt{\xi^{2}+(y/x_{0})^{2}}\rangle}\,\,\,\,\,\,\,\,\mathtt{from}\,\,\frac{\langle\rho\rangle}{3P}
\end{align}
 \label{eq:avxirh02resc} \end{subequations}

where, in the last sub-equation, P stands for pressure and $v^{2}$
has been replaced by $3P/\rho$. The plot of $R_{esc}$ averaged in
three ways of Eq.~(\ref{eq:avxirh02resc}) \, are displayed in Fig.
~\ref{fig:resc2xirh1} for a flat universe, but the general trend
for closed and open is also the same:$R_{esc}$ increases to a maximum
value, and then decreases slowly with the expansion of the universe.
But there are drastic differences in the values of the last average
relative to the first two; e.g., for $\Omega_{\gamma\, eq}=\Omega_{\nu\, eq}=0.5$
that is shown, we find that the maximum of $\left(\frac{R_{esc}}{\xi R_{H_{0}}}\right)_{1max}=0.757$
occurs at $\xi=0.847$, that of $\left(\frac{R_{esc}}{\xi R_{H_{0}}}\right)_{2max}=0.477$
occurs at $\xi=0.922$, and $\left(\frac{R_{esc}}{\xi R_{H_{0}}}\right)_{3max}=1.074$
at $\xi=1.632$. These different values indicate a spectrum of possible
length scales. We may combine Eqs. ~\ref{eq:rhr02rh} and ~\ref{eq:xirh02resc}
to write

\begin{equation}
\frac{R_{H}}{R_{esc}}=\sqrt{\frac{f_{esc}(\xi)}{f_{\nu}(\xi)}}\label{eq:rh2resc}
\end{equation}

\begin{figure}[h]

\centering{}\scalebox{.9}{\includegraphics{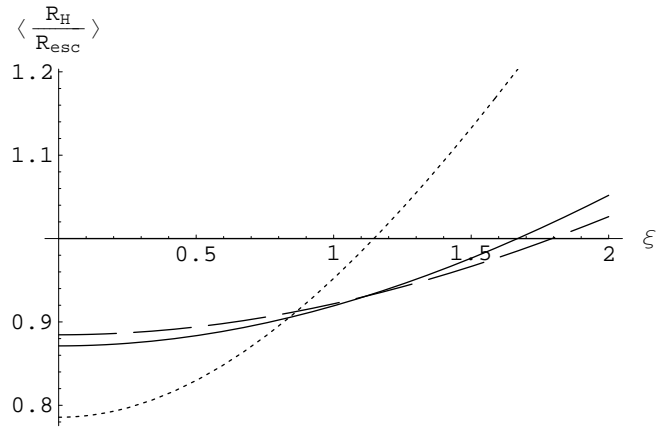}} \caption{The
ratio of $R_{H}/R_{esc}$ for the three averages.The third average
enters the horizon the latest. \label{fig:rh2resc1}}
\end{figure}

This relation, displayed in Fig ~\ref{fig:rh2resc1}, shows that
the horizon crossing at $R_{H}/R_{esc}=1$ occurs at different times
for different averages. $R_{J}$, proportional to $R_{esc}$, will
also show similar behaviour.

The smallest of all these length scales, $R_{esc}$, enters the horizon
the earliest, while the largest, $R_{J}$, enters later. It appears
that the time lag between $R_{J}$ entering horizon at the onset of
Jeans instabilities, the virialization of the cosmic neutrino inside
a smaller $R_{V}$ that entered the horizon earlier, and even the
Keplerization of the gas in an even smaller length $R_{K}$ cannot
be great.

We can also calculate the netrino mass within $R_{esc}$ viz., $M_{esc}=\frac{4\pi}{3}\rho_{\nu}R_{esc}^{3}$
that simplifies to $\frac{M_{esc}}{R_{H_{0}}}=\frac{(y/x_{0})^{3}\xi^{2}}{\sqrt{\Omega_{\nu\, rest\,0}}(\xi^{2}+(y/x_{0})^{2})^{7/4}}$.
Here also, as in the previous section, the averaging over the momentum
of $M_{esc}$ can be done in many different ways. Below, we give three
which we think are useful:

\begin{equation}
\left\langle \frac{M_{esc}}{R_{H_{0}}}\right\rangle _{1}=\frac{0.5}{\sqrt{\Omega_{\nu\, rest\,0}}}\left\langle \frac{(y/x_{0})^{3}\xi^{2}}{(\xi^{2}+(y/x_{0})^{2})^{7/4}}\right\rangle \label{eq:mesc2rh01}
\end{equation}

\begin{equation}
\left\langle \frac{M_{esc}}{R_{H_{0}}}\right\rangle _{2}=\frac{0.5}{\sqrt{\Omega_{\nu\, rest\,0}}}\left\langle \frac{(y/x_{0})^{2}}{\xi^{2}+(y/x_{0})^{2}}\right\rangle \left\langle \frac{(y/x_{0})\xi^{2}}{(\xi^{2}+(y/x_{0})^{2})^{3/4}}\right\rangle ^{1/2}\label{eq:mesc2rh02}
\end{equation}

\begin{equation}
\left\langle \frac{M_{esc}}{R_{H_{0}}}\right\rangle _{3}=\frac{0.5}{\sqrt{\Omega_{\nu\, rest\,0}}}\left\langle \frac{\xi^{2}+(y/x_{0})^{2}}{(y/x_{0})^{2}}\right\rangle ^{-1}\left\langle \frac{(y/x_{0})\xi^{2}}{(\xi^{2}+(y/x_{0})^{2})^{3/4}}\right\rangle ^{1/2}\label{eq:mesc2rh03}
\end{equation}

\begin{figure}[h]

\centering{}\scalebox{.9}{\includegraphics{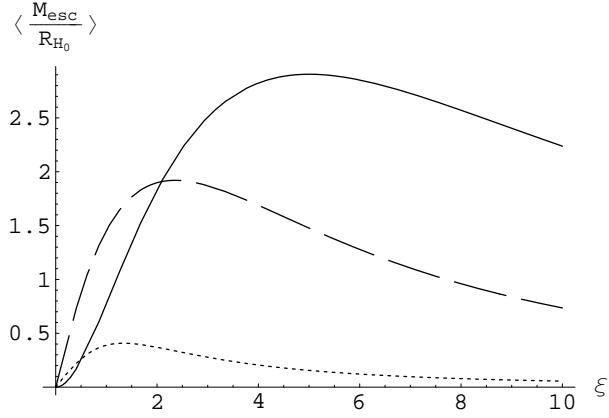}} \caption{The
distribution of Mass contained within escape radius calculated by
different averaging process given in equations
~\ref{eq:mesc2rh01}, ~\ref{eq:mesc2rh02} and
~\ref{eq:mesc2rh03}.The dashed curve represents the value
calculated by the 1st equation, the dotted is that of 2nd equation
and normal curve is from the third one. \label{fig:mesc2rh01}}
\end{figure}

A plot of $(M_{esc}/R_{H_{0}})$ is shown in Fig. ~\ref{fig:mesc2rh01}
for three different averages. Although the nature of the curves are
same, the peaks are seen to occur at different positions.

\section{Virial Equation and Moments}

Moments of the virialized quantities have played very important role
in revealing the features of large astrophysical objects. Some of
these methods are used here to describe the large scale neutrino structures
that should be the precursors of the galactic and the supercluster
scales. The formation of large-scale structures proceeds with the
gravitational collapse of clouds of matter composed of baryons, cold
dark matter and hot dark matter whose major constituent is the neutrino.
Much work has been done to investigate the evolution of size and eccentricity
of spherical and spheroidal / ellipsoidal clusters \cite{Chandra-ellipsoidal,Chandra-elbert,Somsundar},
and and also the relavant masses. In the majority of these works,
Virial theorems / equations and their moments of various orders have
been used. In this section, various moments of the Jeans mass are
calculated using using the ideas of Virial moments. The results are
compared with the values presented in the previous section.

In a system of N particles, gravitational forces tend to pull the
system together and the stellar velocities tend to make it fly apart.
It is possible to relate kinetic and potential energy of a system
through the change of its moment of inertia. In a steady-state system,
these tendencies are balanced, which is expressed quantitatively through
the Virial Theorem. A system that is not in balance will tend to move
towards its virialized state. The Scalar Virial Theorem tells us that
the average kinetic and potential energy must be in balance. The tensor
Virial Theorem tells us that the kinetic and potential energy must
be in balance in each separate direction. The scalar Virial Theorem
is useful for estimating global average properties, such as total
mass, escape velocity and relaxation time, while the tensor Virial
Theorem is useful for relating shapes of systems to their kinematics,
e.g. the flatness of elliptical galaxies to their rotational speed.

The Virial Equations of the various orders are, in fact, no more than
the moments of the relevant hydrodynamical equations. The scalar Virial
Equation for a system is given by

\begin{equation}
\frac{1}{2}\frac{d^{2}I}{dt^{2}}=2K+U
\end{equation}

where the moment of inertia about the origin $I=\frac{1}{2}\intop\rho r^{2}dV$,
the kinetic energy is $K=\frac{1}{2}\frac{d^{2}}{dt^{2}}\intop_{V}\rho v^{2}dV$
and the potential energy is $U=\frac{\rho_{c}}{2}\intop_{V}\frac{\rho}{r}dV$,
where $\rho_{c}$is the core density. Thus the Virial Equation of
the 1st order is given by

\begin{equation}
\frac{1}{2}\frac{d^{2}}{dt^{2}}\intop_{V}\rho r^{2}dV=\frac{1}{2}\frac{d^{2}}{dt^{2}}\intop_{V}\rho v^{2}dV+\frac{\rho_{c}}{2}\intop_{V}\frac{\rho}{r}dV
\end{equation}

The Virial Equation of the 2nd order is given by just multiplying
the integrand by r before integrating:

\begin{equation}
\frac{1}{2}\frac{d^{2}}{dt^{2}}\intop_{V}\rho r^{3}dV=\frac{1}{2}\frac{d^{2}}{dt^{2}}\intop_{V}\rho v^{2}rdV+\frac{\rho_{c}}{2}\intop_{V}\rho dV
\end{equation}

Similarly, the higher order equation may be written. For the steady
state, 2K + U = 0. This gives the virial radius of the spherical cluster:

\begin{equation}
R_{V}^{2}=\frac{3}{4\pi}\left(\frac{v^{2}}{\rho}\right)
\end{equation}

Writing $\rho$ and v in terms of x = ma = m/T and y = pa = p/T, as
discussed in Introduction section, we have

\begin{equation}
R_{V}=Constant.\frac{y}{\left(y^{2}+x^{2}\right)^{7/4}}\left(\frac{x}{m}\right)^{2}
\end{equation}

and hence the expectation value of this virial radius is given by

\begin{equation}
<R_{V}>=Constant.\intop_{0}^{\infty}dy\frac{y^{3}}{\left(e^{y}+1\right)\left(y^{2}+x^{2}\right)^{7/4}}\left(\frac{x}{m}\right)^{2}
\end{equation}

\begin{figure}
\includegraphics[scale=0.8]{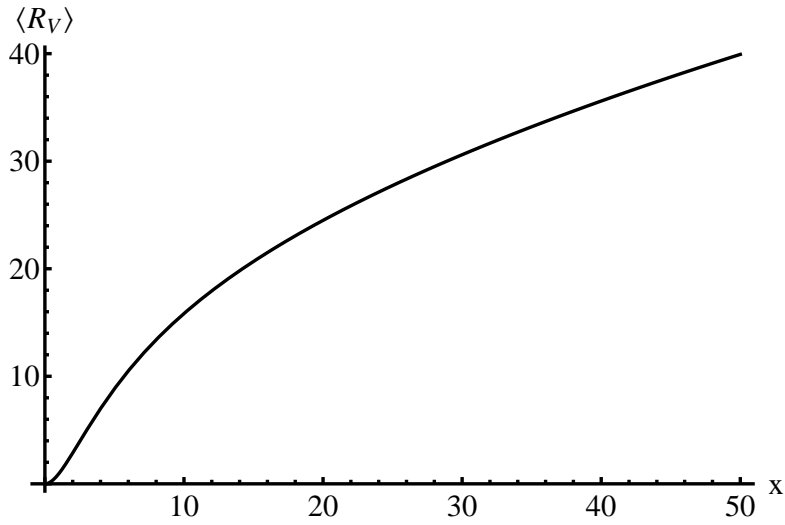}

\caption{\label{fig:Rv_vs_x}Evolution of Virial Radius $<R_{V}>$ of a cluster.}
\end{figure}

Its evolution has been shown in \ref{fig:Rv_vs_x}.

\begin{figure}
\includegraphics[scale=0.9]{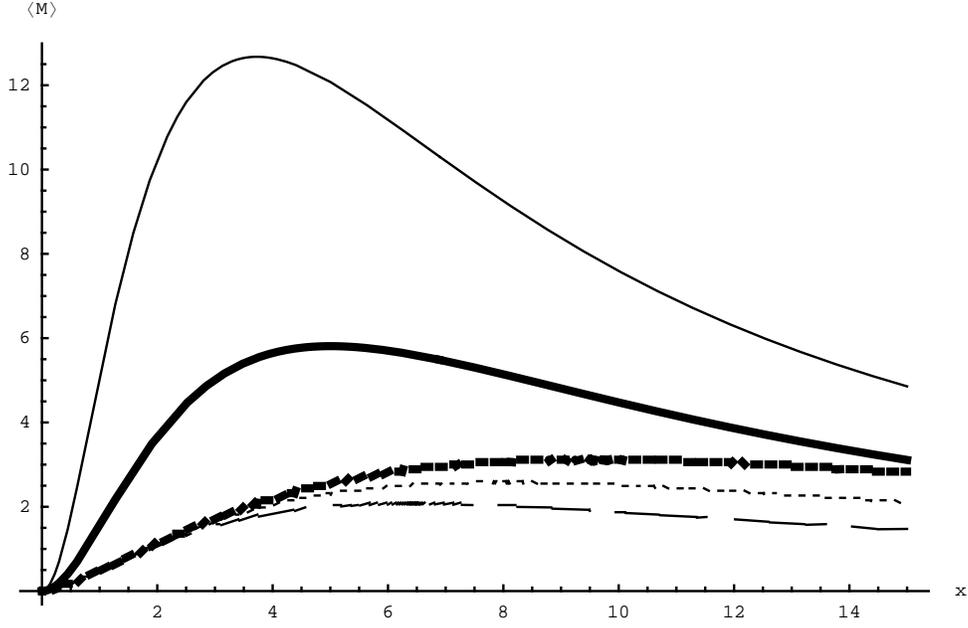}

\caption{\label{fig:JeansMasses_virialized}The average cluster masses (Jeans
and Virial) against x. The normal curve is for virial mass, thick
continuous curve for Jeans mass, the normal dashed curve for the Jeans
mass calculated from the 1st moment of the mass and normal dotted
and thick dotted are those from 2nd and 3rd moments respectively.}
\end{figure}

The Virial Mass of the cluster contained within this radius may be
calculated as $<M_{V}>=constant.<R_{V}>^{3}<\rho>$ . Its variation
with x is as expected (\ref{fig:JeansMasses_virialized}), but peaks
at x = 3.73 in contrast to the position of peaks of Jeans masses calculated
in section \ref{sec:-Neutrino-Jeans-mass}, where the peaks have occurred
at x = 1.9, 2.1, 4.2 and 5.0. \cite{Prem,Bond}

\subsection{Virial moments of Jeans mass}

As stated earlier, in the virial method, we take the moments of the
equation of motion. These equations obviously involve the moments
of the distribution of density, pressure, velocity, gravitational
potential, etc. Here we are taking the spatial moments of various
orders of Jeans mass given by \ref{jeansmass}: The 1st moment of
Jeans mass is given by

\begin{eqnarray}
\left\langle M_{J}R_{V}\right\rangle  & = & constant.\intop_{0}^{\infty}dy\frac{y^{2}}{\left(e^{y}+1\right)}(M_{J}R_{V})\nonumber \\
 & = & Constant.\intop_{0}^{\infty}dy\frac{y^{6}x^{4}}{\left(e^{y}+1\right)\left(y^{2}+x^{2}\right)^{5/2}}
\end{eqnarray}

\begin{figure}
\includegraphics{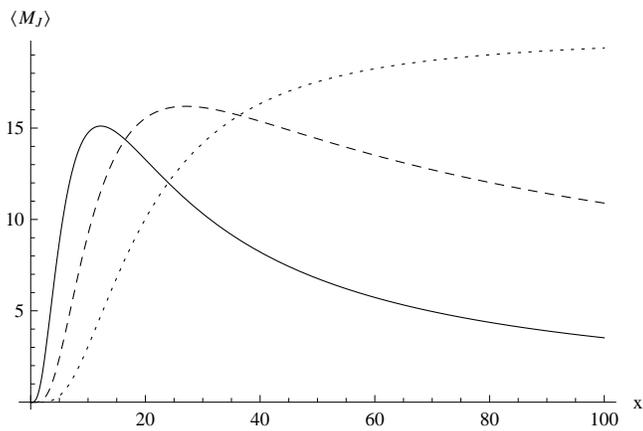}

\caption{\label{fig:VirializedMoments_MJ}Virialized moments of the first three
orders of Jeans masses. The first moment peaks at x \textasciitilde{}
$m_{\nu}$ / T = 6.48. The second moment peaks at x = 8.01 and the
3rd one at 9.58.The solid, dashed and dotted lines are for 1st, 2nd
and 3rd order moments respectively. They are appropriately scaled
to make of equal heights for the comparision purpose.}
\end{figure}

The 2nd and 3rd moments may be written in similar fashion. The variations
of the 1st, 2nd and 3rd moments with x have been shown in \ref{fig:VirializedMoments_MJ}.The
1st and the 2nd are found to peak and at x = 12.2 and 27.02 respectively,
but the 3rd moment appears to plateau off from x \textasciitilde{}
100 to large values of x. Jeans masses may be calculated by dividing
these moments by $<R_{V}>$, $<R_{V}^{2}>$ and $<R_{V}^{3}>$ respectively.
The variation of these masses with x has been shown in \ref{fig:JeansMasses_virialized}
(For comparison $<M_{J}>$ and $<M_{V}>$ have also been plotted).
They have peaked at x = 6.48, 8.01 and 9.58. Following table gives
the corresponding temperatures at which the peaks of the Jeans masses
occur for a few typical neutrino masses:

\begin{table}
\caption{\label{tab:Temp4VirializedJeansMass}Temperatures at which the peaks
of the Jeans masses occur for different neutrino masses.}

\begin{tabular}{|c|c|c|c|c|}
\hline
\multirow{2}{*}{ $M_{J}$ from} & \multirow{2}{*}{x for $M_{J}$ peaks} & \multicolumn{3}{c}{Temp. (K) corresponding to x for }\tabularnewline
\cline{3-5}
 &  & $m_{\nu}$= 1 eV & $m_{\nu}$= 0.2 eV & $m_{\nu}$= 0.01 eV\tabularnewline
\hline
$\frac{\left\langle \rho v^{2}\right\rangle }{\left\langle \rho\right\rangle ^{2}}$ & 1.9 & 6100 & 1220 & 61\tabularnewline
\hline
$\frac{\left\langle \rho\right\rangle }{\left\langle K_{J}\right\rangle ^{3}}$ & 2.1 & 5519 & 1104 & 55\tabularnewline
\hline
$<R_{V}>^{3}<\rho>$ & 3.73 & 3107 & 621 & 31\tabularnewline
\hline
$M_{J}(y_{rms})$ & 4.2 & 2760 & 552 & 28\tabularnewline
\hline
$\left\langle M_{J}(y)\right\rangle $ & 5 & 2318 & 464 & 23\tabularnewline
\hline
1st moment of $M_{J}$ & 6.48 & 1789 & 358 & 18\tabularnewline
\hline
2nd moment of $M_{J}$ & 8.01 & 1447 & 289 & 14\tabularnewline
\hline
3rd moment of $M_{J}$ & 9.58 & 1210 & 242 & 12\tabularnewline
\hline
\end{tabular}
\end{table}

From the above analysis, it is seen that the large-scale structures
of neutrinos of different mass and random velocity distribution can
form at different neutrino temperatures, corresponding to different
time. The earliest peak that occurs at x = 1.9 corresponds to the
time when the neutrino temperature was 1220 K for the 0.2 eV neutrino.
Similarly, the latest peak occurs at x = 9.58 corresponding to a temperature
of 242 K. In between these two values, it is found that the Jeans
mass peaks at a number of different x. Thus it can be interpreted
to mean that a distribution of neutrino structures of different masses
and of different ages should be in existence. Typical masses of these
structures range from $6\times10^{19}$ to $4.5\times10^{20}M_{\bigodot}$.

\section{\protect\bigskip{}
 Free Streaming}

It is well known that free streaming of collisionless particles wipe
out any structures that form. In this context, there are other two
scales that are relevant. The first is the particle horizon $R_{P}$
occurring at the distance to which light would travel since the big
bang. The comoving particle horizon can be written as
\begin{equation}
\frac{R_{P}}{\xi R_{H_{eq}}}=\int_{0}^{\xi}\frac{d\xi^{\prime}}{\sqrt{f_{\nu}(\xi^{\prime})}}.\label{eq:rp2xirh}
\end{equation}

No massive particle can cover a distance larger than this. The other
is the freestreaming length $R_{F}$, the distance a particle travels
since the big bang; it is given by an expression similar to Eq. ~(\ref{eq:rp2xirh}),
but including the particle speed $v$ inside the integral:

\begin{equation}
\frac{R_{F}}{\xi R_{H_{eq}}}=\int_{0}^{\xi}\frac{vd\xi^{\prime}}{\sqrt{f_{\nu}(\xi^{\prime})}}\label{eq:rf2xirh}
\end{equation}

$=\frac{(\frac{y}{x_{eq}})}{\sqrt{1-\Omega_{eq}}}\int_{y/x_{eq}}^{\sqrt{\xi^{2}+\left(y/x_{eq}\right)^{2}}}\frac{dz}{\sqrt{\left(z^{2}-(\frac{y}{x_{eq}})^{2}\right)\left(z^{2}+\frac{\Omega_{\nu\, rest\, eq}}{1-\Omega_{eq}}z+\frac{\Omega_{\gamma\, eq}}{1-\Omega_{eq}}-\frac{\Omega_{\nu\, rest\, eq}}{1-\Omega_{eq}}\right)}}$

Any scale that is smaller than $R_{F}$ will be wiped out by free
streaming. The integral Eq. ~(\ref{eq:rf2xirh}) in our case, for
different values of $\Omega_{\gamma\, eq}$, can be written as Jacobian
elliptic functions \cite{Byrd}

\begin{equation}
\left(\frac{R_{F}}{\xi R_{H_{0}}}\right)_{flat}=CF(\phi/k)\label{eq:rf2xirf-flat}
\end{equation}

\begin{figure}[h]

\centering{}\scalebox{.9}{\includegraphics{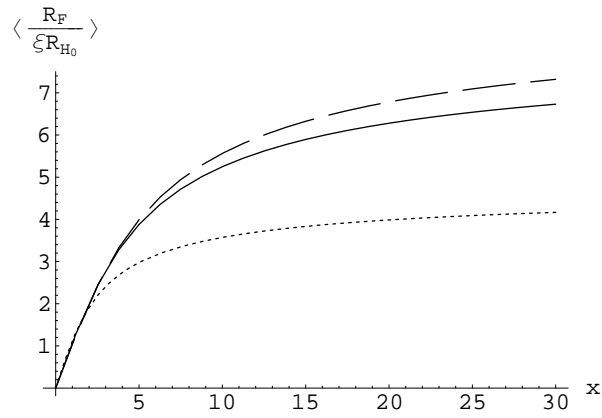}} \caption{The
ratio of $R_{F}/\xi R_{H_{0}}$ for the three averages for the flat
case. \label{fig:rf2xirh01}}
\end{figure}

Considering the flat case where $\Omega_{\gamma\, eq}=0.5$, the solution
is $C^{2}=\frac{8(y/x_{eq})^{2}}{1+\frac{3\zeta(3)}{4\zeta(4)}\frac{y}{x_{eq}}\left(\frac{T_{\nu}}{T_{\gamma}}\right)^{4}}$,
$\sin^{2}\phi=\frac{\sqrt{x_{eq}^{2}+y^{2}}-y}{\sqrt{x_{eq}^{2}+y^{2}}+y}$,
and $k^{2}=\frac{\sqrt{x_{eq}^{2}+y^{2}}-x_{eq}}{\sqrt{x_{eq}^{2}+y^{2}}+x_{eq}}$.
A plot of $(R_{F}/\xi R_{H_{eq}})_{flat}$ is shown in Fig. - ~\ref{fig:rf2xirh01}
for three different mean values of y. This free streaming length is
seen to saturate to the values of 4.193, 3.899 and 2.555 as $x\rightarrow\infty$
, when evaluated with $y_{rms}$, $y_{mean}$ and $y_{rhms}$ respectively.

For non-flat universe, $R_{F}/\xi R_{H_{eq}}$ is given by Eqs. ~(\ref{eq:rf2xirh})
and ~(\ref{eq:rf2xirf-flat}) with $C^{2}=\frac{(y/x_{eq})^{2}}{\Omega_{\gamma\, eq}}\sqrt{1+(y/x_{eq})^{2}}$,
$\tan^{2}(\phi/2)=\sqrt{\frac{(\sqrt{x_{eq}^{2}+y^{2}}-y)}{(\sqrt{x_{eq}^{2}+y^{2}}+y)}}\frac{(\sqrt{x_{eq}^{2}+y^{2}}-y)}{(\sqrt{x_{eq}^{2}+y^{2}}+y)}$
and $2k^{2}=1+(y/x_{eq})^{2}\left(\frac{1-\Omega_{eq}}{\Omega_{eq}}\right)\sqrt{1+(y/x_{eq})^{2}}$.
Plots of $R_{F}/\xi R_{H_{eq}}$ for $\Omega_{\gamma\, eq}=0.45$,
are found to saturate to 2.883, 2.113 and 1.690, for the three means
of y; and in the closed universe of $\Omega_{\gamma\, eq}=0.55$,
the free streaming length saturates to 2.683,\ 1.908 and 1.495 for
the respective means of y. The nature of the curves are similar to
that of flat ones.

\section{\protect\bigskip{}
 Discussion and conclusion}

From above analysis, it is seen that the large-scale structures of
neutrinos of different mass and random velocity distribution can form
at different neutrino temperatures, corresponding to different time.
The earliest peak that occurs at x = 1.9 corresponds to the time when
the neutrino temperature was 1220 K for the 0.2 eV neutrino. Similarly,
the latest peak occurs at x = 9.58 corresponding to a temperature
of 242 K. In between these two values, it is found that the Jeans
mass peaks at a number of different x and hence it can be interpreted
to mean that a distribution of neutrino structures of different masses
and of different ages should be in existence. Typical masses of these
structures range from $6\times10^{19}\,\, to\,\,4.5\times10^{20}M_{\bigodot}$.
These are in the order of very large super-clusters which formed first
in the hot dark matter scenario. But we should note that as the Universe
cools down, and x becomes large, the scale of these structures will
come down because the leading term in the Jeans mass, Eq. (\ref{jeansmass})
goes as $x^{-1/2}$. Thus for 1 eV neutrino at the present temperature
of 1.96 K ($\sim10^{-4}eV$), the Jeans mass comes down to the galactic
scale. Accordingly, the large superclusters that formed initially
would have fragmented to the galactic size.

Thus the massive neutrino can play a very important role in the formation
of structures. These particles decouple from the rest of matter at
a very early time when the temperature was ~1 MeV. Since then they
have been cooling down independently without interacting with other
matter except gravitation. So the evolution of this component can
be considered to occur separately. We have done the Jeans analysis
taking into account only neutrino. The results show the possibility
of different sizes of neutrino structures forming at different time.
In particular, two distinct sizes make their appearance: the smaller
ones which could form at relatively higher temperature of $T\sim m/2$
when the neutrinos were still relativistic, and the ten time bigger
ones that form when $T\sim m/5$. As these neutrino structures have
a mass of order of $m_{pl}^{3}/m^{2}$, these are indeed very large,
in the scale of large super-clusters. The Jeans mass at lower temperature
is of the order of galaxy. In this work we have considered the neutrino
as a totally independent component of the Universe that interacts
with nothing except gravitation with each other. In reality, these
neutrino interact gravitationally with the rest of the matter. So
in any volume, there is more matter than just neutrino that is trying
to bring about a gravitational collapse. Hence the actual scale at
which the neutrino could begin to collapse should be smaller than
the value we have derived.

\end{document}